\documentclass[12pt]{JHEP3}

\usepackage{epsfig}
\epsfclipon

\def\Dsl{\hbox{/\kern-.6700em\it D}} 
\def\dsl{\hbox{/\kern-.5300em$\partial$}}

\newcommand{\sfrac}[2]{{\textstyle\frac{#1}{#2}}}
\def\eq{\begin{equation}}
\def\eeq{\end{equation}}
\def\eqa{\begin{eqnarray}}
\def\eeqa{\end{eqnarray}}
\def\nn{\nonumber}

\def\bd{\begin{displaymath}}
\def\ed{\end{diplaymath}}

\def\Box{ {\,\lower 0.9pt\vbox{\hrule\hbox{\vrule height0.2cm \hskip 0.2cm \vrule height 0.2cm }\hrule}\,}}
\def\lsim{{\ \lower-1.2pt\vbox{\hbox{\rlap{$<$}\lower5pt\vbox{\hbox{$\sim$}}}}\ }}
\def\gsim{{\ \lower-1.2pt\vbox{\hbox{\rlap{$>$}\lower5pt\vbox{\hbox{$\sim$}}}}\ }}

\def\pref#1{(\ref{#1})}

\def\ssubsubsection#1{\vspace{3mm} \noindent \textbf{#1} \\ \vspace{-3mm} \\ \noindent}

\def\cL{{\cal L}}

\def\Dsl{\hbox{/\kern-.6700em\it D}} 
\def\dsl{\hbox{/\kern-.5300em$\partial$}}

\def\nn{\nonumber}

\def\beginvector{\left( \begin{array}{c} }
\def\endvector{\end{array} \right)}
\def\endignore{}
\def\ignore#1\endignore{}

\def\Sph2{{\mathcal S}^2}

\title{Quintessentially Flat Scalar Potentials}
\author{C.P.~Burgess, P. Grenier and D. Hoover
\\

Physics Department, McGill University,
                3600 University Street,\\
                Montr\'eal, Qu\'ebec, Canada, H3A 2T8.}

\abstract{Both inflationary and quintessence cosmologies require
scalar fields which roll very slowly over cosmological time
scales, and so typically demand extremely flat potentials.
Sufficiently flat potentials are notoriously difficult to obtain
from realistic theories of microscopic physics, and this poses a
naturalness problem for both types of cosmologies. We propose a
brane-world-based microscopic mechanism for generating scalar
potentials which can naturally be flat enough for both types of
cosmological applications. The scalars of interest are
higher-dimensional bulk pseudo-Goldstone bosons whose scale of
symmetry breaking is exponentially suppressed in the
higher-dimensional theory by the separation between various
branes. The light scalars appear in the effective 4D theory as
pseudo-Goldstone bosons. Since naturalness problems are more
severe for quintessence models, motivated by our construction we
explore in more detail the possibilities for using
pseudo-Goldstone bosons to build quintessence models. Depending on
how the cosmological constant problem is solved, these models
typically imply the universe is now entering a matter-dominated
oscillatory phase for which the equation of state parameter $w =
p/\rho$ oscillates between $w = 1$ and $w = -1$.}

\preprint{McGill-03/16}


\begin{document}

\section{Introduction}
Perhaps the most interesting consequence of the recent spate of
cosmological measurements is the accumulation of evidence
suggesting the Universe has passed through no less than {\it two}
independent periods of acceleration during that part of its
history to which we have observational access. The first of these
periods is the early inflationary period \cite{Inflation}, whose
simplest predictions for the temperature fluctuations of the
cosmic microwave background (CMB) radiation appear to describe
very successfully what is seen \cite{WMAPInflation}. The big
surprise of the past decade is the discovery that the present
epoch also appears to be a period of incipient inflation, as
indicated by both CMB measurements \cite{WMAPparams} and supernova
surveys \cite{SN}.

Both of these epochs can be described by the slow roll of a scalar
field since they are both defined by the condition that the
universal expansion accelerates, and this in turn requires the
dominant contribution to the energy density, $\rho$, to have
sufficiently negative pressure: $p < - \rho/3$.\footnote{We assume
here the universe to be spatially flat, $k=0$.} If the dominant
energy is due to a rolling homogeneous scalar field, then its
pressure-to-energy ratio is related to the fraction, $r = K/V$, of
the scalar field's kinetic and potential energies according to
$p/\rho = (r-1)/(r+1)$. This shows that acceleration is possible
only if the scalar field is presently potential-energy dominated:
$K \lsim V$. For inflation the corresponding energy density is
typically chosen to be $\rho \approx V \lsim (10^{15} \;
\hbox{GeV})^4$, while applications to the present epoch --- which
we generically refer to as `quintessence' models
\cite{Quintessence} --- instead require $\rho \approx V \sim
(10^{-3} \; \hbox{eV})^4$.

Both of these applications of slow-roll scalar fields run into
difficulties because of the flatness of the potential which they
require. The problem is the notorious difficulty in obtaining very
flat potentials from realistic theories of microscopic physics
\cite{Naturalness}. Our purpose in this paper is to propose a new
mechanism for obtaining extremely flat potentials from within a
brane-world picture \cite{braneworld,BIQ}. In the model we
propose, the scalar of interest is a pseudo-Goldstone boson
\cite{pGB,PhysRep} for an approximate symmetry (more about which
below) which is explicitly broken, but whose breaking requires the
presence of more than one brane as well as of a massive field
living in the bulk between the branes. This combination ensures
that the low-energy effective potential is suppressed by the
amplitude, ${\cal A}$, for the massive particle to propagate from
one brane to another, which is exponentially small in the
inter-brane separation, $a$, in units of the massive-particle
Compton wavelength, $M^{-1}$: ${\cal A} \sim \exp(- M a)$.

The paper is organized in the following way. In the next section
we outline how flat the scalar potentials must be for cosmological
applications to inflation and quintessence, and summarize the
naturalness problems which one encounters trying to obtain
potentials this flat. This section also very briefly reviews what
it means for the scalar be a pseudo-Goldstone boson (pGB), and why
this can help with the naturalness issues. Since this section is
not particularly new (see
refs.~\cite{FHSW,CormierHolman,pGBcosmology} for applications of
pseudo-Goldstone bosons to cosmology), the professionals will want
to skip directly to the next section, $\S 3$, where we describe
our brane-world model, and show why it can give such flat
potentials. At low energies the scalar model we produce is a
pseudo-Goldstone boson, and so motivated by this in $\S 4$ we
build an explicit quintessence model in order to show a detailed
example of a successful cosmology using pseudo-Goldstone bosons,
updating the earlier models of refs.~\cite{FHSW,CormierHolman}. In
this section we also re-examine the viability of these models in
the light of recent WMAP measurements, and identify
potentially-observable differences between this kind of cosmology
and other proposals. Finally, our conclusions are summarized in
section $\S 5$.

\section{Slow-Rolling Scalars and Cosmology}
In this section we have two goals. First, we review the general
constraints which cosmological applications require of
slowly-rolling scalar fields, to see what kinds of hierarchies of
scale a successful cosmology requires of an underlying theory.
Then we examine pseudo-Goldstone bosons in particular, and ask how
large the corresponding heirarchies are related to the
corresponding energy scales for the various types of symmetry
breaking. In this second section we identify two cases, which
differ in whether or not the largest symmetry-breaking effects
arise in the scalar kinetic energies or in the scalar potential.

Since the arguments in this section are relatively standard,
experts should feel free to skip directly to section $\S 3$.

\subsection{Constraints Required by Cosmologically Slow Rolls}
A problem with applications of rolling scalar fields to both
inflation and to quintessence cosmologies arises because they each
require an inordinately flat potential. We here summarize these
constraints subject to very mild assumptions.

In general, a scalar roll is only slow enough to neglect its
kinetic energy if the slow-roll parameters \cite{LL} $\epsilon =
\sfrac12 \, (M_p V'/V)^2$ and $\eta = M_P^2 V''/V$ are much
smaller than unity. (Here $M_p \approx 10^{18}$ GeV is the 4D
Planck mass and the prime denotes differentiation with respect to
the canonically-normalized scalar fields.) To see what this
requires suppose the scalar action has the generic form
\eq \label{pGBgenform}
   - {{\cal L} \over \sqrt{-g}} = \frac{f^2}{2} \, (\partial
   \varphi)^2 + \mu^4 \, v(\varphi) \,,
\eeq
where $0 \le \varphi(x) \le 2 \pi$ is a dimensionless field and
$f$ and $\mu$ are constants having dimensions of mass (in units
for which $\hbar = c = 1$). If we suppose that $v(\varphi)$ and
all of its derivatives are $O(1)$, then $\epsilon \sim \eta \sim
M_p^2/f^2$ which shows that we must require $f \gg M_p$, in which
case the scalar mass, $m \sim \mu^2/f$, must be much smaller than
the Hubble scale, $H = \Bigl( {\rho/3M_p^2} \Bigr)^{1/2} \sim
\mu^2/M_p$.

\ssubsubsection{Inflation}
In order for such a rolling scalar in an early inflationary period
to properly describe the amplitude of CMB temperature fluctuations
requires the combination $\delta^2 = (1/150\pi^2) (V/M_p^4
\epsilon)$ must satisfy $\delta \approx 2 \times 10^{-5}$. Using
the conditions $\epsilon \sim M_p^2/f^2$ and $H \sim \mu^2/M_p$
just described, implies $\mu/M_p \sim 0.03 \, \sqrt{M_p/f}$.
Together with the observational constraint \cite{WMAPInflation}
$\epsilon  < 0.03$, we find the requirement $M_p /f \lsim 0.2$ and
$\mu/M_p \lsim 0.006$.

\pagebreak

\ssubsubsection{Quintessence}
On the other hand, the situation is even worse if the scalar is to
describe today's Universal acceleration, since such a scalar must
satisfy $\mu \sim 10^{-3}$ eV. This, with the slow-roll condition
$f \gsim M_p$, leads to $\mu/f \lsim \mu/M_p \sim 10^{-30}$ and
the incredibly small scalar mass $m \sim \mu^2/f \lsim \mu^2/M_p
\sim 10^{-33}$ eV.

\subsection{Naturalness Issues}
It is notoriously difficult to get very flat scalar potentials
from realistic microscopic physics without fine-tuning, and this
difficulty comes in two parts. First one must ask: {\it How do the
small ratios $\mu/f$ and $\mu/M_p$ arise within the microscopic
theory as a combination of microscopic parameters?} Given that
such a small ratio is predicted by the microscopic physics, one
must then ask: {\it How does it remain small as one integrates out
all the physics between these microscopic scales and the
cosmological scales at which it is measured?}

Of these, the second problem is the more serious, the more so the
lower $\mu$ is required to be. It is a problem because a particle
of mass $M$, which interacts with the scalar with order-unity
couplings, typically shifts $\mu$ by an amount $\delta \mu \propto
M$ when it is integrated out, which can be unacceptably large if
$M \gsim \mu$. There are two symmetries which are known to be able
to help with this problem, in that they can ensure that particles
of mass $M$ do not produce corrections as large as $\delta \mu
\sim M$. The two symmetries are: (1) supersymmetry, for which
bose-fermi cancellations ensure $\delta \mu \lsim M_s$, where
$M_s$ is the typical mass splitting within a supermultiplet; (2)
Goldstone symmetries, for which the scalar transforms
inhomogeneously according to $\delta \varphi = \epsilon [1 +
F(\phi)]$, where $\epsilon$ is the transformation parameter and
the potentially nonlinear function $F$ satisfies $F(0) = 0$.

This second type of symmetry arises only if the scalar in question
is a Goldstone boson for a spontaneously broken global symmetry,
and it ensures $v(\varphi)$ must be completely independent of
$\varphi$. $v(\varphi)$ can be nontrivial if the global symmetry
is only approximate, in which case corrections to $\mu$ are
systematically suppressed by whatever small symmetry-breaking
parameter makes the symmetry a good approximation. In this case
the scalar $\varphi$ is known as a pseudo-Goldstone boson
\cite{pGB,PhysRep}.

\subsection{Pseudo-Goldstone Bosons}
The scales $\mu$ and $f$ are related to the scales of symmetry
breaking in the underlying microscopic theory. Once set there they
naturally remain small as successive scales are integrated out to
obtain an effective theory at very low energies. These low-energy
corrections remain small precisely because the scalar $\varphi$ is
a pseudo-Goldstone boson, and so corrections to $\mu$ are
protected by the nonlinearly-realized $G$ symmetry.

A lower limit to the amount of this suppression can be inferred
completely within the low-energy theory by power-counting within
it the size of loop-generated symmetry-breaking corrections
\cite{PhysRep}. To this end consider a system of $N$ scalars,
$\varphi^a$, where we choose to rescale the spacetime metric to go
to the Einstein Frame, for which the graviton kinetic term takes
the canonical Einstein-Hilbert form. The scalar part of the
lagrangian density which involves the fewest derivatives may
always be written
\eq \label{scalaraction}
    {\cL_s \over \sqrt{-g}} = - \, V(\varphi)
    - \, \frac12 \, G_{ab}(\varphi) \, g^{\mu\nu}
    \partial_\mu \varphi^a \; \partial_\nu \varphi^b \,.
\eeq
The symmetric tensor $G_{ab}$ may be interpreted as a metric on
the scalar-field `target' space. For a single scalar field
$G_{ab}$ may be set to unity by an appropriate field redefinition,
and so in this case it is the scalar potential, $V(\varphi)$,
which determines all of the physics. A similar choice, $G_{ab} =
\delta_{ab}$ is {\em not} possible if $N \ge 2$, however, unless
the target space happens to be flat. In order to avoid missing
physics associated with $G_{ab}$ we consider models below
involving two or more pseudo-Goldstone scalars.

When the scalars $\varphi^a$ are Goldstone bosons the functions
$V$ and $G_{ab}$ are strongly restricted by symmetry conditions.
These imply $V$ must be a constant, and for the symmetry-breaking
pattern $G \to H$, $G_{ab}$ must be a metric on the coset space
$G/H$ whose isometries include the symmetry group $G$. This
usually determines $G_{ab}$ up to a few constants, and often
completely determines it up to overall normalization and field
redefinitions \cite{CCWZ,PhysRep}.

For example, for the symmetry-breaking pattern $SO(3) \to SO(2)$
there are then two Goldstone bosons, $(\varphi^1, \varphi^2) =
(\theta, \phi)$, which parameterize the coset space $SO(3)/SO(2)$,
which in this case is the two-sphere, $S_2$. Here we use standard
spherical-polar coordinates, $0 \le \theta < \pi$ and $0 \le \phi
\le 2 \pi$, on $S_2$. The $SO(3)$ transformations amount to the
rotations of this sphere about its centre, if $S_2$ is embedded
into Euclidean three-dimensional space. The condition that the
action be invariant under $SO(3)$ transformations then requires
$V(\theta,\phi)$ to be constant, and $G_{ab}(\theta,\varphi)$ to
be the standard rotationally-invariant -- `round' -- metric on the
2-sphere:
\eq \label{E:Round} G_{ab} \, \partial_\mu \varphi^a \partial^\mu
\varphi^b = f^2 \; \Bigl(
\partial_\mu \theta \, \partial^\mu \theta + \sin^2 \theta \,
\partial_\mu \phi \, \partial^\mu \phi \Bigr) . \eeq
$f$ is a dimensionful constant whose size indicates the scale of
spontaneous symmetry breaking.

Our interest here is in pseudo-Goldstone bosons, for which the
global symmetry $G$ is only approximate in the sense that the
effective energy scale, $\mu$, associated with the explicit
breaking of the symmetry is much smaller than the scale, $f$, of
its spontaneous breaking. (We have already seen that this
effective scale need not be simply related to the microscopic
scales of the microscopic theory.) In this case $V$ need no longer
be independent of $\varphi^a$ and $G_{ab}$ need not be a
$G$-invariant metric, although deviations from these limits should
be small if the scale, $\mu$, is much smaller than the scale, $f$,
of spontaneous symmetry breaking.

In the limit $f \gg \mu$ on dimensional grounds we expect the
generic corrections to $V$ to be of order $\mu^4$ and corrections
to $G_{ab}$ to be of order $\mu^2/f^2$. If the asymmetric terms
are initially this size, they automatically remain so after being
renormalized by quantum corrections within the low-energy theory.
As is briefly summarized in the appendix, these orders of
magnitude can differ in supersymmetric theories, if $\mu$ is
larger than the supersymmetry breaking scale.

For instance, for the $SO(3)/SO(2)$ example, suppose the $SO(3)$
symmetry is explicitly broken but the $SO(2)$ symmetry associated
with shifting $\phi$ is not. Then examples of the kinds of new
terms one might expect at low energies might be
\eqa \label{SO3breakingterms}
    V(\varphi) &=& a + \sum_{n \ge 1} b_n \cos(n \theta) \\
    G_{ab} \; d\varphi^a \, d\varphi^b &=& f^2 \left[ d\theta^2 +
    G(\theta) \, d\phi^2 + \dots \right] \, , \nonumber \\
    \hbox{where} \qquad G(\theta) &=&
    \sin^2\theta + \sum_{n \ge 2} c_n \, \sin^2 (n \theta) \,,
\eeqa
where the sums run over integer values. The above dimension
counting then argues that while $a$ need not be suppressed by
$\mu$, we expect $b_n \lsim \mu^4$ and $c_n \lsim \mu^2/f^2$.

In applications it is usual to neglect the corrections to $G_{ab}$
and keep only the scalar potential which is induced by explicit
symmetry breaking. This is usually justified because the
symmetry-breaking potential always dominates at low energies
because there is no symmetry-invariant potential with which to
compete. For instance, in cosmological applications Hubble damping
inevitably slows the scalar motion, making the potential a more
and more important influence on the scalar roll. The same is not
true for the corrections to the target-space metric, $G_{ab}$,
since these are always at most of order $\mu^2/f^2$ relative to
the $G$-invariant metric of the symmetry limit.

\ssubsubsection{Global Symmetries and Gravity}
The power-counting statements made above assume that quantum
corrections respect the theory's underlying $G$ invariance.
Unfortunately, there is an important kind of quantum correction
which may not do so for any global symmetry, and this represents a
potential obstacle to using a pseudo-Goldstone symmetry to keep
the corrections to $\mu$ small.

The problem comes from gravitational quantum corrections, which
are believed not to respect global symmetries, for instance due to
the virtual appearance and disappearance of black holes (which the
`no-hair' theorems ensure cannot carry global-symmetry charges).
Estimates \cite{GravSym} of the amount of symmetry breaking which
this induces in a low-energy 4D effective theory predict that the
symmetry-breaking interactions are suppressed by powers of
$f^2/M_p^2$. This can represent an important renormalization to
$\mu$ precisely for the case of cosmologically slowly-rolling
fields, for which we've seen $f \gsim M_p$.

One must keep in mind that these estimates of non-perturbative
quantum-gravity effects carry the caveat that they assume many
things about the properties of quantum gravity at high energies,
and so may not properly capture how things work once this
high-energy physics is better understood. In particular, as
pointed out in ref.~\cite{GravSym}, these symmetry-breaking
estimates can change dramatically if the effective theory becomes
higher dimensional at scales $M_c \ll M_p$, as is the case in the
brane-world models we describe below. Of course, how small a
quantum gravity correction may be tolerated depends very much on
how flat a scalar potential is desired, making all of these issues
much more pressing for present-epoch quintessence models than they
are for inflation.

In what follows we proceed under the assumption that this, or a
similar mechanism, ensures that high-energy quantum-gravity
effects do not destroy the effectiveness of the pseudo-Goldstone
boson mechanism in protecting the flatness of the scalar potential
to the accuracy required for cosmology.

\section{Flat Scalar Potentials from the Brane World}
Although pseudo-Goldstone bosons can have naturally flat
potentials if $\mu \ll f$, they do not in themselves explain why
$\mu$ should be so small. An understanding of this must come from
a more microscopic theory. In this section we describe a
brane-world model within which such small scales can arise for
pseudo-Goldstone boson potentials. Brane models are natural to
examine from this point of view, because in many situations they
have given new insights on how small quantities can arise in
low-energy low-energy physics \cite{natbrane}.

The idea behind our construction is to make a model having a $G =
U_A(1) \times U_B(1)$ global symmetry which is spontaneously
broken by the {\it vev}, $v$, of a bulk scalar field $\Phi$. The
symmetry is also broken explicitly by the couplings of a second
bulk scalar field $\Psi$ (having a large mass $M$) to various
brane fields $\chi_i$. In particular, the model is designed so
that there is more than one brane (say two of them) and only the
$U_A(1)$ symmetry is broken by the $\Psi$ couplings to the first
brane, and only the $U_B(1)$ symmetry is broken by the $\Psi$
couplings to the second brane which is displaced a distance $a$
away from the first brane. Once this is arranged, functional
integration over $\Psi$ and the brane modes generates a nontrivial
scalar potential for the would-be goldstone mode in $\Phi$, which
is suppressed by the amplitude ${\cal A} \propto \exp(- M a)$ for
the field $\Psi$ to propagate from one brane to the other. The
logic of this construction is reminiscent of brane-based
supersymmetry breaking mechanisms, for which each brane preserves
some supersymmetries but where all supersymmetries are broken by
at least one brane \cite{braneSSB}.

\subsection{The Higher-Dimensional Toy Model}
Consider, then, a model containing the complex scalar bulk fields
$\Phi$ and $\Psi$, and complex brane fields $\chi_i, i = 1,2$,
whose action is $S = S_B + S_{b1} + S_{b2}$, with
$(4+n)$-dimensional bulk action
\eqa \label{Baction}
    S_B &=& -\int d^4x \, d^ny \; \Bigl[ (\partial \Psi)^*
    (\partial \Psi) + (\partial \Phi)^* (\partial \Phi) +
    V(\Phi,\Psi) \Bigr] \nn\\
    \hbox{where} \qquad V(\Phi,\Psi) &=&
    M^2 \Psi^* \Psi + \frac{\lambda}{2} (\Phi^*\Phi - v^2)^2
     \, ,
\eeqa
and 4-dimensional brane actions
\eqa \label{baction}
    S_{b1} &=& - \int_{y=y_1} d^4x \; \Bigl[ (\partial \chi_1)^* (\partial
    \chi_1) + m_1^2 \, \chi_1^* \chi_1 + \sfrac12\, [ (g_1 \Phi + h_1 \Psi) \,
    \chi_1^2 + \hbox{c.c.} ] \Bigr] \nn\\
    S_{b2} &=& - \int_{y=y_2} d^4x \; \Bigl[ (\partial \chi_2)^* (\partial
    \chi_2) + m_2^2 \, \chi_2^* \chi_2 + \sfrac12\, [ (g_2 \Phi + h_2 \Psi^*) \,
    \chi_2^2 + \hbox{c.c.} ] \Bigr] \, .
\eeqa
We assume all of the couplings, $\lambda, g_i$ and $h_i$ to be
real and nonzero, but sufficiently small to permit a perturbative
analysis of the model. We imagine the branes to be parallel
3-branes which are situated at the points $y = y_i$ within the $n$
compact transverse dimensions. We take the size of all of these
dimensions to be of the same order, $r$, making the
compactification scale (Kaluza-Klein masses) of order $M_c \sim
1/r$.

By construction, the model enjoys a global $G = U(1) \times
\tilde{U}(1)$ symmetry under which each of the bulk scalars rotate
independently: $\Phi \to e^{i \omega} \Phi$ and $\Psi \to e^{i
\tilde\omega} \Psi$. Other perturbative couplings could also be
permitted in the bulk scalar potential without substantially
changing our conclusions, provided they also respect this
symmetry.

The brane couplings, on the other hand, each explicitly break this
symmetry down to a single $U(1)$. The bulk-brane couplings at
brane 1 preserve the subgroup $U_A(1)$ under which $\Phi \to
e^{i\omega_A} \Phi$, $\Psi \to e^{i\omega_A} \Psi$ and $\chi_1 \to
e^{-i\omega_A/2} \chi_1$. Similarly the couplings on brane 2 only
preserve the subgroup $U_B(1)$ under which $\Phi \to e^{i\omega_B}
\Phi$, $\Psi \to e^{-i\omega_B} \Psi$ and $\chi_2 \to
e^{-i\omega_B/2} \chi_2$. Taken together, both branes completely
break the symmetry group $U(1) \times \tilde{U}(1)$. Notice also
that it is only the field $\Psi$ which transforms differently
under $U_A(1)$ and $U_B(1)$, and so both the brane and bulk
actions would preserve one of the $U(1)$'s if the field $\Psi$
were everywhere set to zero.

The spectrum of this model is easy to understand in the limit $h_i
\to 0$, in which case the brane couplings also preserve the $G$
symmetry. Then the nonzero expectation value $\langle \Phi \rangle
= v$ spontaneously breaks the $U(1)$ symmetry, while leaving the
$\tilde{U}(1)$ unbroken. In the absence of the branes, therefore,
the bulk theory would consist of a mass-$M$ complex field $\Psi$
plus the two real mass eigenstates coming from $\Phi$. One of the
$\Phi$ mass eigenstates would in this case be a massless Goldstone
boson, $\varphi = \arg\Phi$, and the other would have a mass of
order $\sqrt\lambda \; v$. We now compute how nonzero $h_i$
couplings on the brane change these conclusions.

\subsection{The Effective 4D Theory}
Since our interest is in the would-be Goldstone boson, we focus on
the low-energy theory below the compactification scale, by
integrating out all of the massive fields on the branes and in the
bulk. In particular, we concentrate on the effective scalar
potential for the would-be Goldstone mode, $\varphi$, in this
low-energy theory. We therefore look for those terms in the scalar
potential which involve the phase of $\Phi$, neglecting also the
Kaluza-Klein tower of compactification modes for this field.

\pagebreak

\ssubsubsection{Integrating out the Brane Modes}
We start by integrating out the brane scalars, with the bulk
fields held fixed. The leading contribution arises at one loop,
leading to a scalar-potential contribution to the effective
$(4+n)$-dimensional bulk theory of the form
\eq \label{bindpot}
    \delta {\cal L}_B = \sum_{i=1}^2 \Delta V_{{\rm eff},i} \;
    \delta^n(y - y_i) \, ,
\eeq
where
\eq
    \Delta V_{{\rm eff},i} = \frac{1}{64\pi^2}\;
    \hbox{Tr}
     \left[ {\cal M}^4_i
    \log \left( \frac{ {\cal M}^2_i }{\mu^2} \right) \right] \, ,
\eeq
and the trace is over the two components of the $\chi_i$ mass
matrix
\eq
    {\cal M}^2_i = \pmatrix{ m_i^2 & \xi_i \cr \xi_i^* & m^2_i
    \cr} \,.
\eeq
Here $\mu$ is an arbitrary renormalization scale, $\xi_1 = g_1 \,
\Phi + h_1 \, \Psi$ and $\xi_2 = g_2 \, \Phi + h_2 \, \Psi^*$.
Evaluating the trace we find
\eqa \label{DVexp}
    \Delta V_{{\rm eff},i} &=& \frac{1}{64\pi^2}
     \left[ (m^2_i + |\xi_i| )^2 \log \left( \frac{m^2_i +
     |\xi_i| }{\mu^2} \right) +
    (m^2_i - |\xi_i| )^2 \log \left( \frac{m^2_i -
     |\xi_i| }{\mu^2} \right) \right] \nn\\
     &\approx& \frac{1}{32\pi^2}
     \left[ m^4_i \log \left( \frac{m^2_i}{\mu^2} \right) +
    |\xi_i|^2 \left( \sfrac32 + \log \left( \frac{m^2_i}{\mu^2} \right)
    \right) + \cdots \right] \, ,
\eeqa
where we assume $m_i$ to be large enough to ensure that $m_i^2 >
|\xi_i|$ for $\Phi \sim v$, and the second, approximate, equality
applies for $|\xi_i| \ll m^2_i$.

\ssubsubsection{Integrating out $\Psi$}
We next integrate out the massive bulk field, $\Psi$, with $\Phi$
temporarily held fixed. The leading result in this case arises at
tree level, corresponding to the elimination of $\Psi$ from the
classical action (supplemented by the effective brane-induced
interaction, eq.~\pref{bindpot}), using its classical field
equation
\eq
    \Bigl( - \Box  + M^2 \Bigr) \Psi_c = -  \sum_{i=1}^2 \delta^n(y-y_i)
    \; \frac{\partial \Delta V_{{\rm eff},i}}{\partial \Psi^*}  \,
    ,\nn
\eeq
which, using eq.~\pref{DVexp}, gives the approximate expression
\eqa
    \Bigl( - \Box  + M^2 \Bigr) \Psi_c
    &\approx& - \,\delta^n(y-y_1) \; \frac{h_1 \, \xi_1}{32\pi^2}
    \left[ \sfrac32 + \log\left( \frac{m^2_1 }{\mu^2} \right) \right] \\
    && \qquad \qquad - \,
     \delta^n(y-y_2) \; \frac{h_2 \, \xi_2^*}{32\pi^2}
    \left[ \sfrac32 + \log\left( \frac{m^2_2 }{\mu^2} \right) \right] \,
    .\nn
\eeqa

Working to leading order in powers of $h_i$ allows us to write
$\xi_i \approx g_i \Phi$ in this last equation, allowing its
solution to be written
\eqa
    \Psi_c(y) &\approx& - \, \frac{1}{32 \pi^2} \, \left\{ h_1\,
    g_1 \, \Phi(y_1) \,  \left[ \sfrac32 +
    \log\left( \frac{m^2_1 }{\mu^2} \right) \right] \, G(y,y_1)
    \right. \nn\\
    && \qquad\qquad \left. + h_2\,
    g_2 \, \Phi^{*}(y_2) \,  \left[ \sfrac32 +
    \log\left( \frac{m^2_2 }{\mu^2} \right) \right] \, G(y,y_2)
    \right\} \, .
\eeqa
Here $G(y,y')$ is the solution to $(- \Box + M^2) \, G(y,y') =
\delta^n(y,y') - 1/\Omega_n$, where $\Omega_n$ denotes the volume
of the $n$ extra dimensions. $G(y,y')$ is given explicitly by the
mode sum
\eq
    G(y,y') =  {\sum_\ell}' \; \frac{ u_\ell(y) \;
    u^*_\ell(y')}{\lambda_\ell} \, ,
\eeq
in terms of the eigenvalues and eigenfunctions satisfying $(- \Box
+ M^2) u_\ell(y) = \lambda_\ell \, u_\ell(y)$. The prime on the
sum indicates the omission of any zero modes, for which
$\lambda_\ell = 0$.

Substitution into the classical action, eqs.~\pref{Baction} and
\pref{baction}, then gives an action of the form $S_{\rm
eff}[\Phi] = S_{\rm inv}[\Phi] + \Delta S[\Phi]$, where $S_{\rm
inv}[\Phi]$ is invariant with respect to $\Phi \to e^{i \omega }
\Phi$, and
\eq
    \Delta S[\Phi] \approx
     - k \int d^4x \; \Bigl[ g_1 \, g_2 \,  h_1\,  h_2\, \Phi(x,y_1) \,
     \Phi(x,y_2) \, G(y_1,y_2) + \hbox{\rm c.c.} \Bigr] + \cdots
    \, .
\eeq
In this last expression the constant $k$ is given explicitly by
\eq
    k = \left( \frac{1}{32\pi^2} \right)^2 \, \left[ \sfrac32 +
    \log \left( \frac{m_1^2}{\mu^2} \right) \right] \left[
    \sfrac32 + \log \left( \frac{m_2^2}{\mu^2} \right) \right] \,
    .
\eeq

\ssubsubsection{Integrating out the $\Phi$ Kaluza-Klein Modes}
The final step is to integrate out the massive Kaluza-Klein modes
for $\Phi$ to obtain the effective four-dimensional action. Since
the Kaluza-Klein zero mode for $\Phi$ is independent of the
extra-dimensional coordinates $y$, to leading order this
corresponds to simply truncating the action using $\Phi(x,y) \to
\Phi(x)$.

Using the information that the invariant part of the potential is
minimized for $\Phi(x) = v_R \, e^{i \varphi(x)} \ne 0$, where
$v_R$ is an appropriately renormalized parameter which differs
from $v$ because of the changes to the invariant part of the
potential (which we do not follow here in detail), we obtain in
this way the following effective action for the would-be Goldstone
mode, $\varphi$:
\eq \label{pGBaction}
    S_{\rm eff}[\varphi] = - \int d^4x \; \Bigl[f^2
    (\partial \varphi)^2 + V(\varphi) \Bigr] \, ,
\eeq
where $f^2 = v^2_R \, \Omega_n$ with $\Omega_n$ as before denoting
the volume of the internal dimensions. The low-energy scalar
potential is given within the above approximations by $V \approx
\mu^4 \, \cos(2 \varphi) + \dots$ (up to an additive constant,
$V_0$, which we may absorb into the renormalization of the
cosmological constant). The constant $\mu$ is given approximately
by
\eq \label{pGBpotential}
    \mu^4 \approx   2\, k \, g_1 \, g_2 \,  h_1\,  h_2\, v^2_R \,
    G(y_1,y_2) \, .
\eeq
Eqs.~\pref{pGBaction} and \pref{pGBpotential} are the main results
of this section.

\subsection{Phenomenological Choices for the Scales}
We see that the higher-dimensional model implies an effective 4D
action for $\varphi$ of the generic form of eq.~\pref{pGBgenform},
with the constants $f$ and $\mu$ given in terms of more
microscopic parameters. Given an internal space for which
$\Omega_n \sim r^n$ and a brane separation $a$, we therefore find
the order of magnitude results
\eq
    f \sim v \, r^{n/2} \, ,
\eeq
and
\eq \label{muresult}
    \mu \sim (g_1 \, g_2 \, h_1 \, h_2)^{1/4} \sqrt{v} \, \left[
    \frac{1}{M} \,
    \left( \frac{M}{a} \right)^{(n-1)/2} \, e^{- Ma} \right]^{1/4}
    \, ,
\eeq
where we take $v_R$ and $v$ to be the same order of magnitude.

For comparison the 4D Planck mass is given by $M_p \sim M_g \,
(M_g \, r)^{n/2}$, where $M_g$ is the higher-dimensional
gravitational scale. Eq.~\pref{muresult} uses the asymptotic form
for the Green's function in the limit $Ma \gg 1$: $G(a) \sim
M^{-1} \, (M/a)^{(n-1)/2} \, \exp(-Ma)$.

The exponential dependence of the heavy-field Green's function is
what allows the scale $\mu$ to be naturally much smaller than $f$
and $M_p$. For example, consider the simplest instance where we
assume $r$ is much larger than all other fundamental length
scales, which we choose to all be of order $M_g$. Taking then $a
\sim r \gg 1/M_g$, we therefore suppose the higher-dimensional
theory to involve only a single scale, $M_g$, and so take $g_i
\sim \hat{g}_i \, M_g^{1 - n/2}$, $h_i \sim \hat{h}_i \, M_g^{1 -
n/2}$, $M \sim M_g$ and $v \sim M_g^{1+n/2}$. This leads to
\eq
    f \sim M_p \sim M_g \Bigl( M_g \, r \Bigr)^{n/2} \qquad \hbox{and}
    \qquad
    \mu \sim \frac{(\hat{g}_1 \, \hat{g}_2 \, \hat{h}_1 \, \hat{h}_2
    )^{1/4}}{(M_g \, r)^{(n-1)/8}} \, M_g \, \exp\left( - \, \frac{M_g
    \, r}{4} \right) \, .
\eeq
This expression shows that the most natural choice for the
higher-dimensional scales implies a large decay constant $f \sim
M_p$, but with the ratio $\mu/f$ exponentially small given even
only a moderately large value for $M_g r \gg 1$.

The exponential dependence on $M_g a$ allows the resulting scale
$\mu$ to easily be small enough even for present-epoch
applications. For instance taking $\hat{g}_i \sim \hat{h}_i \sim
1$ and $M_g r$ to be only slightly larger than the minimum size
required to solve the electroweak hierarchy problem \cite{BIQ},
$M_g r \sim 200$, $n = 6$ and $M_g \sim 10^{11}$ GeV, we have $\mu
\sim 10^{-3}$ eV.

Applications to inflation are also possible provided it can be
ensured that $f \gg M_p$. For instance this might be arranged in
one of the above scenarios if $M_\Phi \sim \sqrt{\lambda} v \sim
M_g$, but with $\lambda \ll M_g^{-n/2}$. In this case the
requirement $\mu \sim 10^{-4} \, M_p$ requires a smaller
microscopic hierarchy. For instance if $n= 6$ then $M_g \sim
10^{15}$ GeV and $M_g r \sim 8$ does the job.

\section{Pseudo-Goldstone Boson Cosmologies}
Given the extremely shallow potentials which are possible with
this mechanism, we next re-examine the cosmology of
pseudo-Goldstone boson models in more detail. Our purpose in so
doing is to reconsider more quantitatively the cosmological
viability of these models in the light of present observations.

We first describe in general the cosmological rolling of several
scalar fields in four dimensions, and then return to the specific
cases where the scalars are pseudo-Goldstone bosons. Our purpose
is to define our notation, and to highlight the features of
generic pGB-based Quintessence models so these may be contrasted
with what obtains for the usual axion-based models
\cite{FHSW,CormierHolman}.

\subsection{General Multi-scalar Equations}
The equations of motion which are obtained by varying the sum of
the Einstein-Hilbert and the scalar action of
eq.~\pref{scalaraction} produce the following equations of motion:
\eqa \label{EOM}
    R_{\mu\nu} + \kappa^2 \Bigl[ G_{ab} \partial_\mu
    \varphi^a \,
    \partial_\nu \varphi^b + V(\varphi) g_{\mu\nu} \Bigr] &=& 0 \nn\\
    g^{\mu\nu} D_\mu \partial_\nu \varphi^a - G^{ab} V_{,b} &=& 0 ,
\eeqa
where $V_{,a} = \partial V/\partial \varphi^a$ and we adopt
Weinberg's curvature conventions \cite{G&C}. The spacetime and
target-space covariant derivative, $D_\mu$, for the scalar field
which appears in eq.~\pref{EOM} is defined by:
\eqa
    D_\mu \partial_\nu \varphi^a &=& \nabla_\mu \partial_\nu
    \varphi^a + \Gamma^a_{bc}(\varphi) \, \partial_\mu \varphi^b
    \partial_\nu \varphi^c
    \nn\\
    &=& \partial_\mu \partial_\nu \varphi^a - \gamma^\lambda_{\mu\nu}
    \partial_\lambda \varphi^a + \Gamma^a_{bc} \, \partial_\mu \varphi^b
    \partial_\nu \varphi^c.
\eeqa
$\gamma^\mu_{\nu\lambda}(x)$ is the usual Christoffel symbol
constructed from the spacetime metric $g_{\mu\nu}(x)$ and
$\Gamma^a_{bc}(\varphi)$ is the Christoffel symbol built from the
target-space metric $G_{ab}(\varphi)$.

For cosmological applications we restrict these equations to a
homogeneous but time-dependent field configuration and a
Friedmann-Robertson-Walker (FRW) spacetime: $\varphi^a =
\varphi^a(t)$, and
\eq
    g_{\mu\nu}dx^\mu dx^\nu = - dt^2 + a^2(t)\, \gamma_{mn}(y) \,
    dy^m dy^n \, ,
\eeq
where $\gamma_{mn}$ is the usual homogeneous metric on the
surfaces of constant $t$, parameterized by $k=0,\pm 1$. With these
choices the equations of motion reduce to:
\eqa \label{phicosm1}
    H^2 = \left( {\dot a \over a} \right)^2 &=& {\rho \over 3
    M_p^2} - {k \over a^2} \nn\\
    {d\over da}\Bigl(\rho \, a^3 \Bigr) &=& -3p \, a^2 \nn\\
    {D \dot \varphi^a \over dt} + 3 H \dot \varphi^a + G^{ab} \,
    {\partial V \over \partial \varphi^b}
    &=& 0 \,,
\eeqa
where
\eqa \label{phicosm2}
    \rho &=& \frac12 \, G_{ab} \, \dot \varphi^a \dot \varphi^b +
    V(\varphi) \nn\\
    p &=& \frac12 \, G_{ab} \, \dot \varphi^a \dot \varphi^b - V(\varphi) \\
    {D \dot \varphi^a \over d t} &=& \ddot \varphi^a +
    \Gamma^a_{bc}(\varphi) \dot \varphi^b \dot \varphi^c . \nn
\eeqa
Geometrically, the vanishing of $D \dot \varphi^a /dt$ is
equivalent to the statement that $\varphi(t)$ is an
affinely-parameterized geodesic of the target-space metric,
$G_{ab}$.

For instance, for the $SO(3) \to SO(2)$ example the
$SO(3)$-invariant metric has the following nonzero Christoffel
symbols:
\eq \label{Connection0}
    \Gamma^\theta_{\phi\phi} = - \sin\theta \cos\theta, \qquad
    \Gamma^\phi_{\phi\theta} = \Gamma^\phi_{\theta\phi} = \cot\theta.
\eeq
The geodesics of this metric are the `great circles',
corresponding to the intersection of the sphere $S_2 =
SO(3)/SO(2)$, with a plane which passes through the circle's
centre.

Once the $SO(3)$ symmetry is explicitly broken (with the $SO(2)$
unbroken), the symmetry-breaking terms of
eq.~\pref{SO3breakingterms} imply changes to the target-space
connection, leading to the more general expressions
\eq
    \Gamma^\theta_{\phi\phi} = - \, \frac{G'}{2}  \,,\qquad
    \Gamma^\phi_{\theta\phi} = \frac{G'}{2\, G} \, ,
\eeq
where $G' = dG/d\theta$ and all other components are unchanged.
These expressions reduce to eqs.~\pref{Connection0} given the
$SO(3)$-invariant choice $G = \sin^2\theta$.

The qualitative behaviour of the solutions to these equations is
easy to state in the case where the initial scalar kinetic energy,
$K_i$, is large compared with its initial potential energy, $V_i$.
In this case the scalar potential is initially negligible and the
scalar moves along the target-space geodesic determined by its
initial position and velocity. As it so moves the scalar
experiences Hubble friction, which causes it to move more and more
slowly along this geodesic with ever-decreasing kinetic energy.
Eventually its kinetic energy is similar in size to its potential
energy, and so the scalar makes a transition into a
potential-dominated regime. At this point the scalar begins to
follow the gradients of the scalar potential, until it eventually
comes to rest at one of the potential's local minima. A slow roll
occurs if this potential-dominated motion is sufficiently slow.
Because for slow scalar motion both the $\ddot\varphi^a$ and the
$\Gamma^a_{bc} \dot\varphi^b \dot\varphi^c$ terms in the scalar
field equation are small, the entire covariant derivative
$D\dot\varphi^a/dt$ may be neglected during the slow roll.

\subsection{Quintessence Cosmologies} \label{S:Cosmology}
We now examine in more detail the implications of these equations
for applications to present-epoch (quintessence) cosmology. We use
for these purposes the $SO(3) \to SO(2)$ pseudo-Goldstone model of
the previous section. Besides verifying that such cosmologies can
be viable, even after the advent of the WMAP measurements, this
exercise is also useful for identifying those features of the
resulting cosmologies which might be used to distinguish them
observationally from other extant proposals.

\begin{figure}
\vskip 0.25in \centerline{\epsfxsize=4.0in\epsfysize=3.0in
\epsfbox{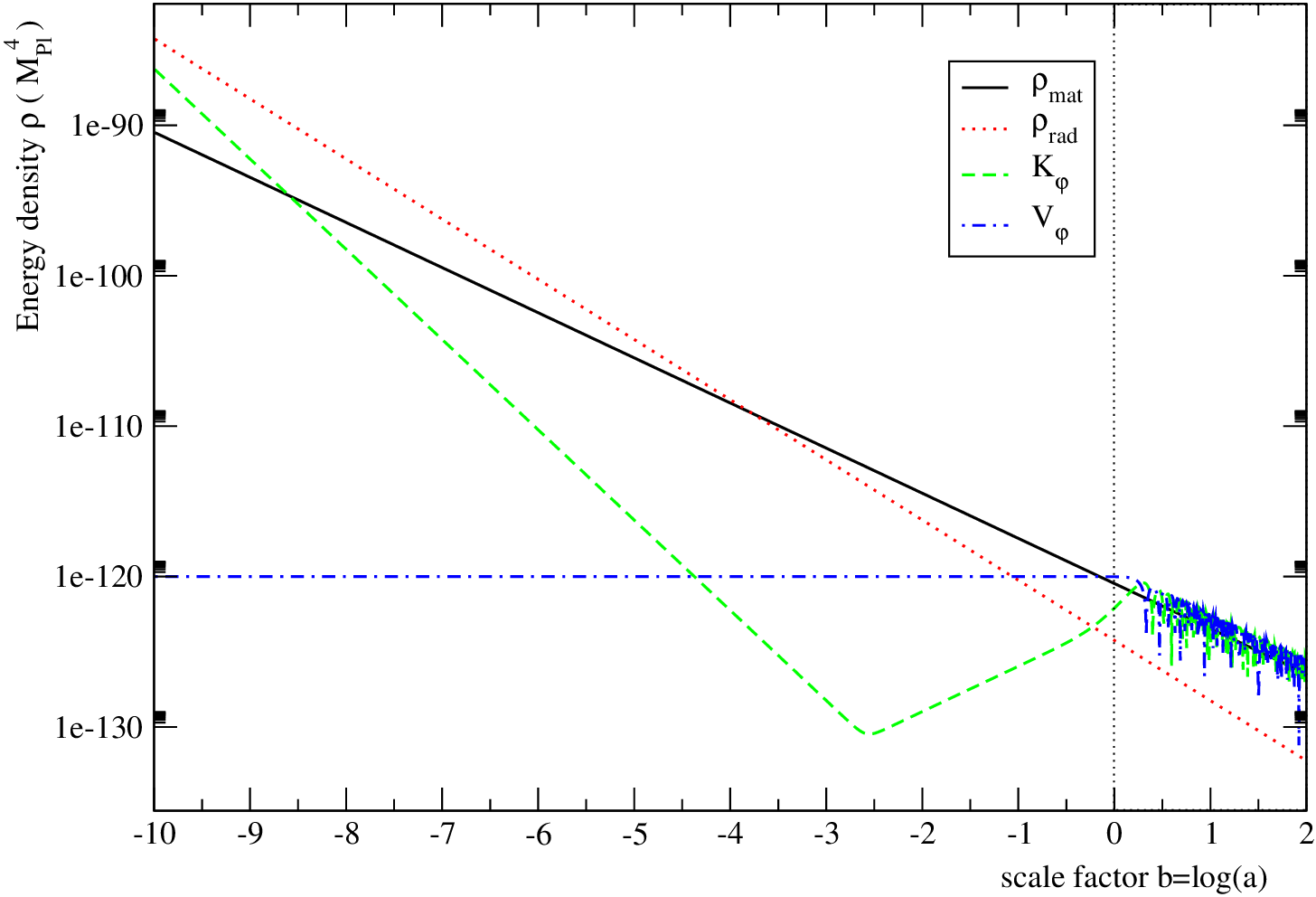}} \caption{Energy density evolution in the
viable quintessence cosmology discussed in the text. Plotted are
the energy density in radiation (red dotted), matter (black line),
total scalar potential (blue dashed-dotted) and total scalar
kinetic (green dashed). Nucleosynthesis occurs at the earliest
epoch shown, and the vertical line indicates the present epoch.}
\label{fig:rhoevolution}
\end{figure}

For concreteness we have explored the model given by
eqs.~\pref{SO3breakingterms}, with the choices $f = M_p$, $a = b_4
= \mu^{4}= \frac12 (10^{-30} M_{p})^{4}$, and $b_2 = b_3 = 0$.
(These arbitrary choices for $b_2$ and $b_3$ are made to arrange
minima for the potential at $\theta = \frac\pi 4$ and
$\frac{3\pi}{4}$, and maxima of the potential at $\theta = 0$,
$\frac\pi 2$ and $\pi$. The main features of the cosmology we
present do not depend on these particular details. $a$ is chosen
to make $V = 0$ at its minima, and this {\it is} important for the
later cosmology. We have no new insights on the cosmological
constant problem in this paper.) Motivated by the simplest
power-counting estimates we also choose $c_n = 0$, although we
return to this choice at the end of this section, where we also
show how our results vary if $c = c_2$ is nonzero.

\begin{figure}
\vskip 0.25in \centerline{\epsfxsize=4.0in\epsfysize=3.0in
\epsfbox{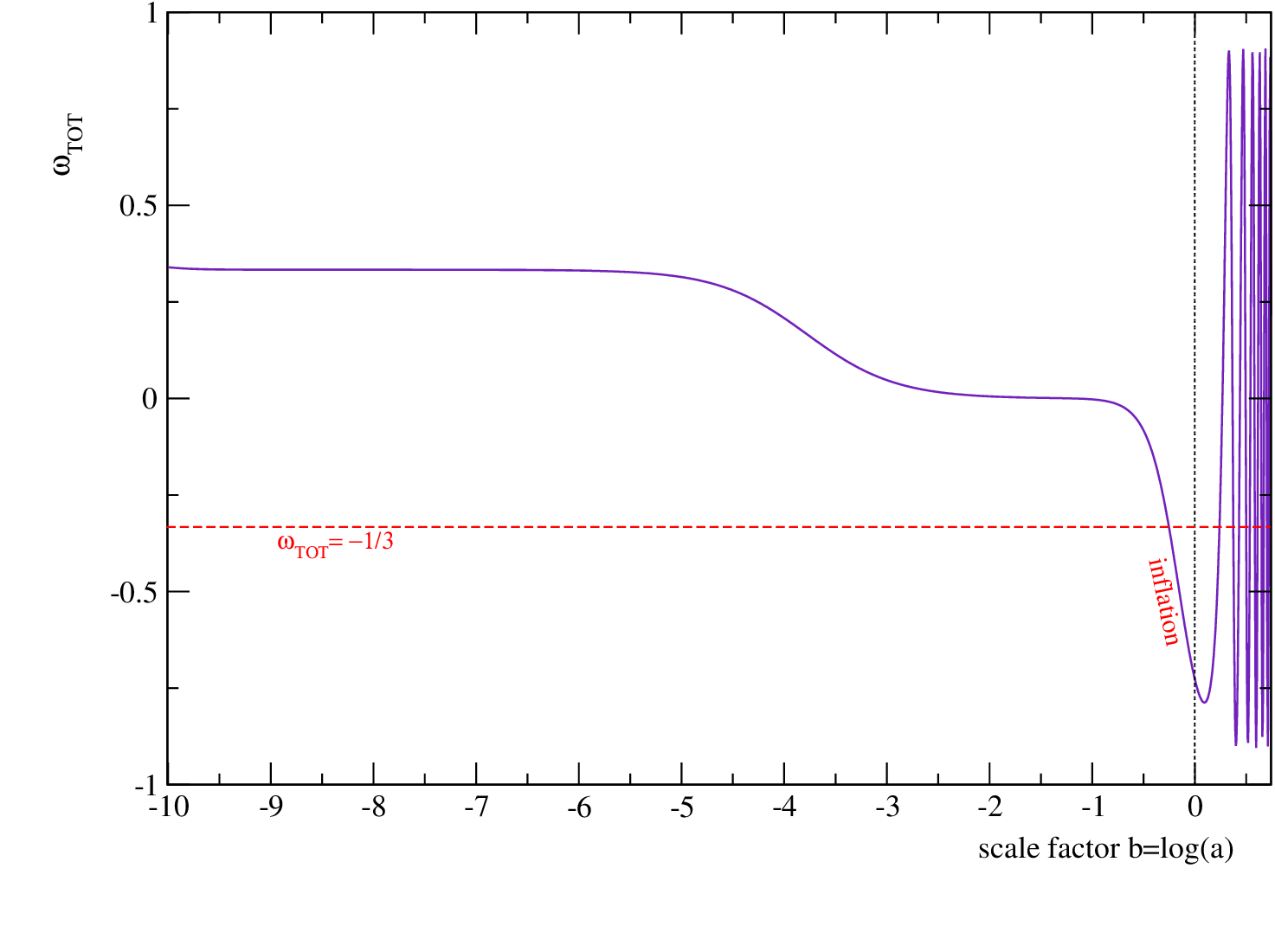}} \caption{Evolution of the equation of state
parameter, $w = p/\rho$, in the cosmology of the previous figures.
The horizontal line marks $w = -1/3$, below which the universe
accelerates.} \label{fig:wtot}
\end{figure}

Fig.~\pref{fig:rhoevolution} shows the results of a numerical
evolution of the field equations for this model, giving the
evolution of the energy density in radiation and matter, as well
as the total kinetic and potential energy density associated with
the scalar field motion. As this figure shows, the scalar fields
in this model are just now entering a period of classical
oscillation about the bottom of their potential, with the total
scalar energy density falling like $1/a^3$ as it is
inter-converted back and forth between kinetic and potential
energy. Although it is at first sight tempting to place the
present epoch during the last period during which $\Omega$ does
not vary appreciably, this option is disfavoured by its
predictions for the equation-of-state parameter $w = p/\rho$, as
may be seen from fig.~\pref{fig:wtot}.

\begin{figure}
\vskip 0.25in \centerline{\epsfxsize=4.0in\epsfysize=3.0in
\epsfbox{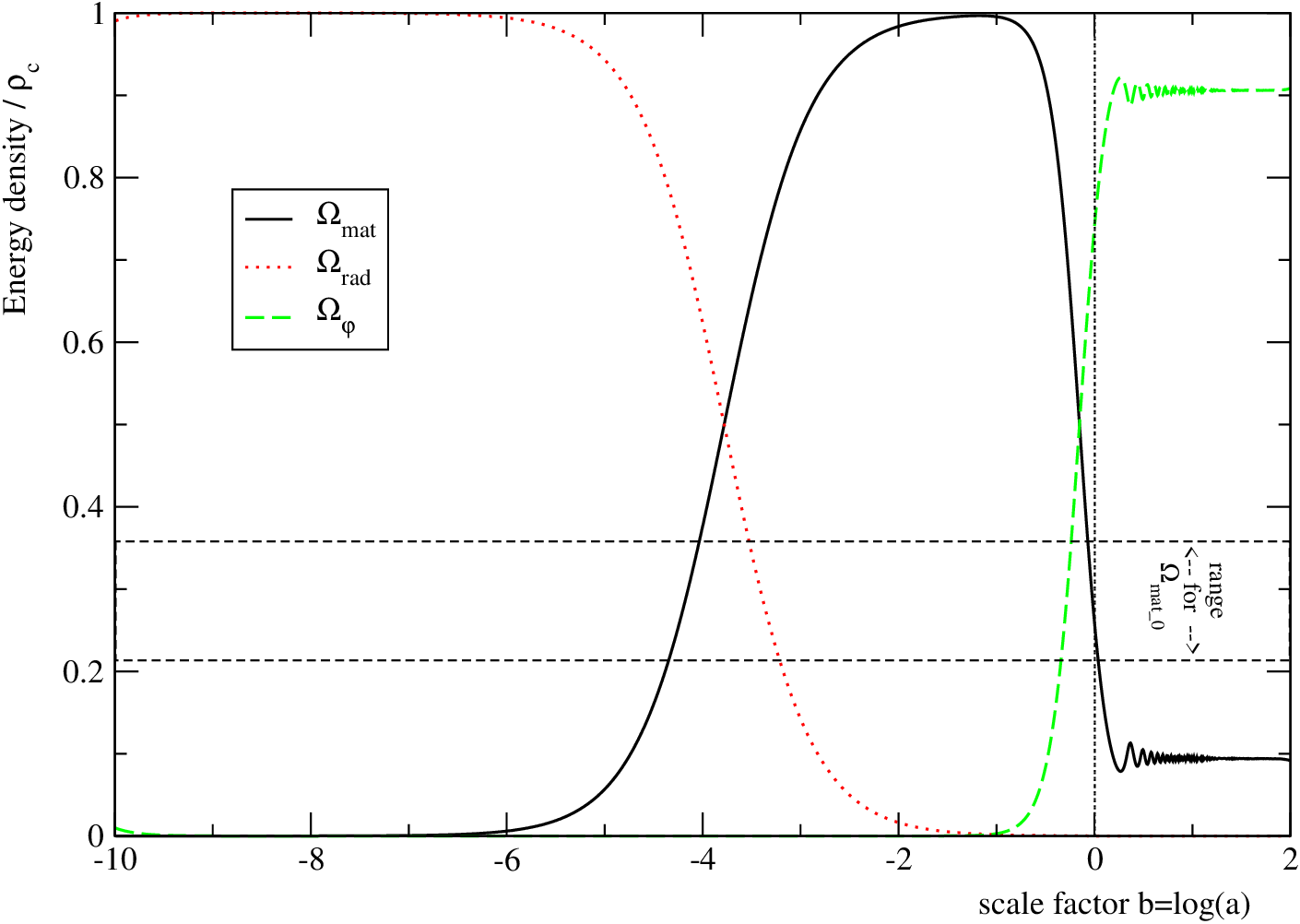}} \caption{The energy density in radiation (red
dotted), matter (black line) and scalars (green dashed) for the
same cosmology as the previous figure, given as a fraction of the
critical density. The horizontal band indicates the
observationally-allowed range for the present-day matter density.}
\label{fig:omega}
\end{figure}

For the cosmology which these figures illustrate, the initial
conditions for the fields $\theta$ and $\phi$ were chosen at the
epoch of nucleosynthesis, with $\theta_0$ near $\pi/2$. The
initial velocities were chosen so that the initial scalar energy
is comparable to the energy in matter and radiation, and since
this is much larger than $V(\theta,\phi)$ this means the scalar
motion is initially dominated by its kinetic energy. Since the
success of standard BBN does not permit the scalar to carry more
than 10\% of the total energy density, we choose the initial
scalar velocities so that $K_\varphi = K_\theta + K_\phi$
saturates this upper bound, with $\dot\theta$ initially zero. Here
$K_\theta = \frac12 \, f^2 \, \dot\theta^2$ and $K_\phi = \frac12
\, f^2 \, G(\theta) \, \dot\phi^2$.

\begin{figure}
\vskip 0.25in \centerline{\epsfxsize=4.0in\epsfysize=3.0in
\epsfbox{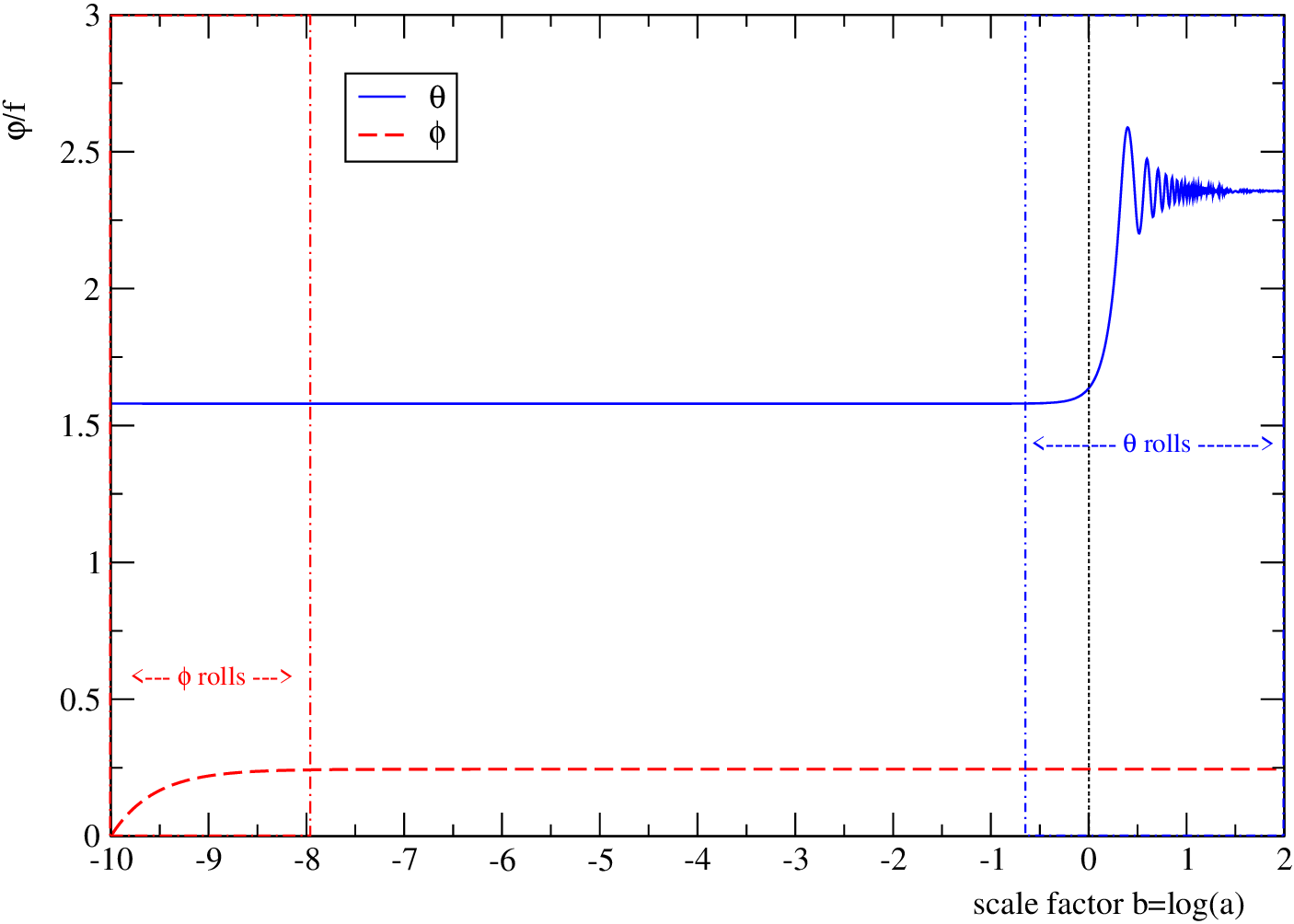}} \caption{Evolution of the two scalar fields,
$\theta$ and $\phi$, with the initial condition $\dot\theta_{BBN}
= 0$.} \label{fig:fields}
\end{figure}

The evolution of the two fields $\theta$ and $\phi$ given these
initial assumptions are then shown in fig.~\pref{fig:fields}. This
figure shows that the $\phi$ evolution is quickly damped by Hubble
friction. Since the scalar potential has maxima for $\theta = 0$
and $\pi/2$ and minima for $\theta = \pi/4$ and $3\pi/4$, the
initial choice $\theta_0 \approx \pi/2$ is close to a maximum.
Once Hubble damping reduces the kinetic energy of the scalars to
close to their potential energy, $\theta$ starts to roll off of
its maximum towards the minimum near $\theta = 3\pi/4$. It is
striking that neither scalar evolves very far, even though their
motion is kinetic-energy dominated for much of the Universe's
history. This feature of the scalar motion may be understood
analytically (see appendix), and is a consequence of the extreme
over-damping due to Hubble friction.

\ssubsubsection{Characteristic Features}
Two features of the scalar motion in this model are generic to
quintessence applications of pseudo-Goldstone boson cosmologies.

\medskip\noindent \textit{Late-Time Oscillatory Cosmology:}
A generic feature of pGB quintessence cosmologies is the late-time
oscillations of the scalar fields about the potential's minimum.
(See, however, ref.~\cite{CormierHolman}, for a model which
differs from most in its late-time consequences.) As is clear from
fig.~\pref{fig:rhoevolution}, although these oscillations are
damped they are not damped faster than the energy density in
matter. As a result the Universe settles down into a comparatively
steady state, for which the relative proportion of energy tied up
in Dark Matter and Dark Energy does not change.

As may be seen from fig~\pref{fig:wtot}, these residual scalar
oscillations may have observational implications because of the
time dependence which they imply for $p/\rho$, and so also for the
acceleration of the Universe. The late-time alternation between
acceleration and deceleration is very different from both the
eternal or temporary inflation predicted by a cosmological
constant or by quintessence based on near-exponential potentials
\cite{Quintessence,ExpPots,AS1,ABRS}, although it is not clear
that this would be observable in the foreseeable future.

\medskip\noindent \textit{Special Initial Conditions:}
A second generic feature of these pGB quintessence models is their
sensitivity to initial conditions. Schematically this sensitivity
arises because a successful cosmology requires the scalar to be
near the maximum of its potential once its kinetic energy becomes
comparable with its potential energy. This ensures the Universe
experiences a sufficiently long period of potential-dominated slow
roll before finally coming to rest at the potential's minimum.

\begin{figure}
\vskip 0.25in \centerline{\epsfxsize=4.0in\epsfysize=3.0in
\epsfbox{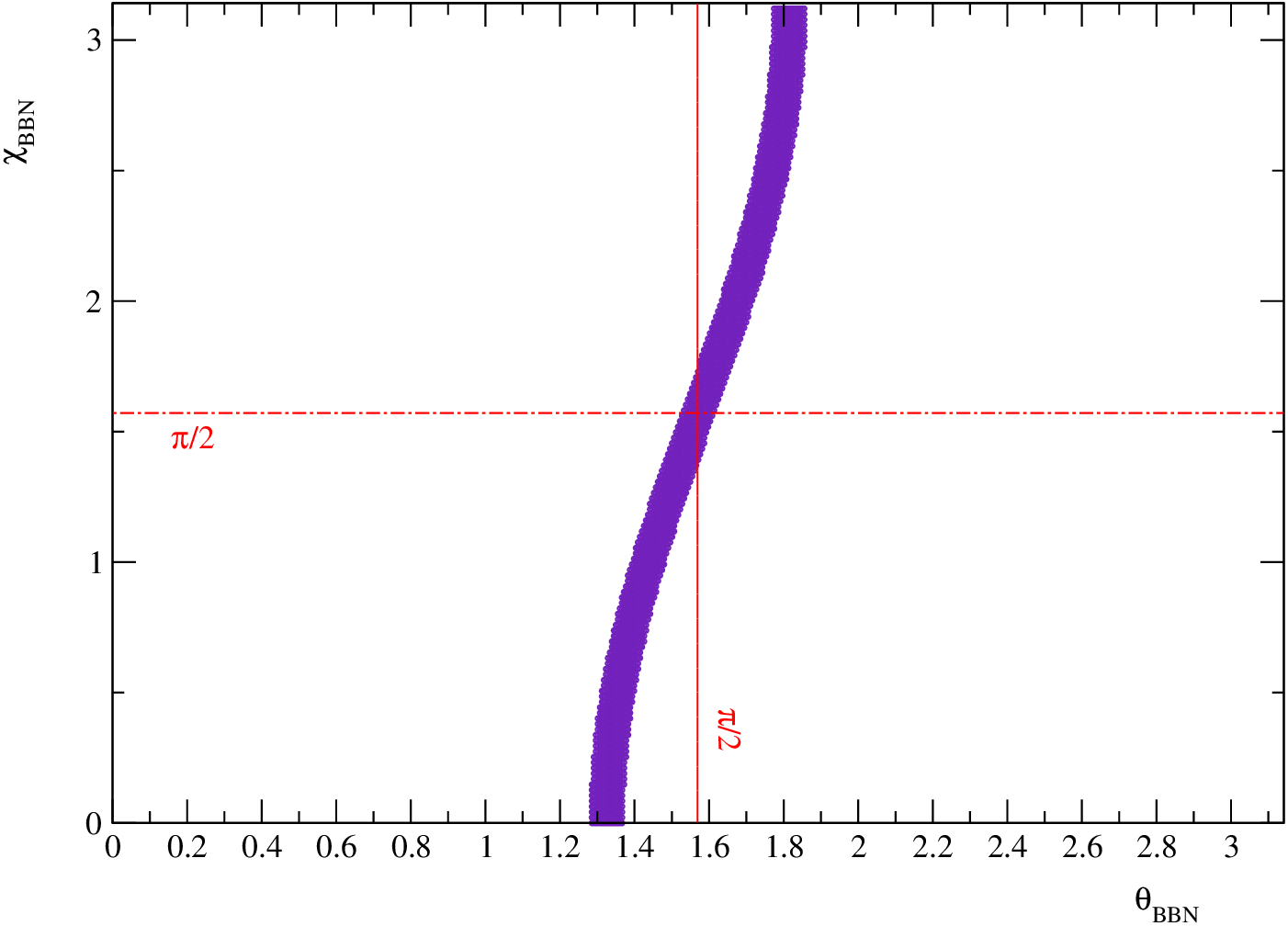}} \caption{Region of initial conditions in the
($\theta_{BBN}-\chi_{BBN}$) plane which give observationally
acceptable values for $\Omega_\varphi$ and $w_{\rm tot}$.}
\label{fig:czero}
\end{figure}

To quantify how broad a class of initial conditions are acceptable
as descriptions of the present-day Universe, we evolved the
cosmology described above for a variety of choices for $\theta_0$
and initial scalar velocities and asked which choices satisfied
the two WMAP constraints \cite{WMAPparams}
\eq
    \frac{K_\varphi}{\rho_{\rm tot}} \approx
    \frac{\rho_\varphi}{\rho_{\rm tot}} = \Omega_\varphi = 0.73 \pm
    0.09 , \qquad
    w = \frac{p_{\rm tot}}{\rho_{\rm tot}} < -0.78 \, ,
\eeq
during the present epoch. Choosing always $K_\varphi = K_\theta +
K_\phi$ to be fixed at 10\% of the total energy density at
nucleosynthesis, we varied the distribution of initial energy
between the two fields $\theta$ and $\phi$ by varying the
parameter $\tan^2\chi = K_\phi/K_\theta$. Fig.~\pref{fig:czero}
shows the region in the initial $\chi-\theta$ plane which satisfy
the two constraints given above. As the figure shows, the allowed
region represents a minor fraction of the area of this plane, but
is also not infinitesimally small.

The shape of the allowed region is easily understood as follows.
It passes through the point $(\chi,\theta) = (\frac{\pi}{2},
\frac{\pi}{2})$ because $\chi = \frac{\pi}{2}$ corresponds to
starting with $\dot\theta = 0$, and $\theta = \frac{\pi}{2}$ is
the maximum of the scalar potential. The corresponding cosmology
simply has $\theta$ remain very nearly at rest at the very top of
the potential from BBN until now. The curve bends away from
$\theta = \frac{\pi}{2}$ as $\chi$ varies because if it starts
with an initial velocity, $\theta$ need not begin at the maximum
at BBN in order to end up there during the present epoch.

\begin{figure}
\vskip 0.25in \centerline{\epsfxsize=4.0in\epsfysize=3.0in
\epsfbox{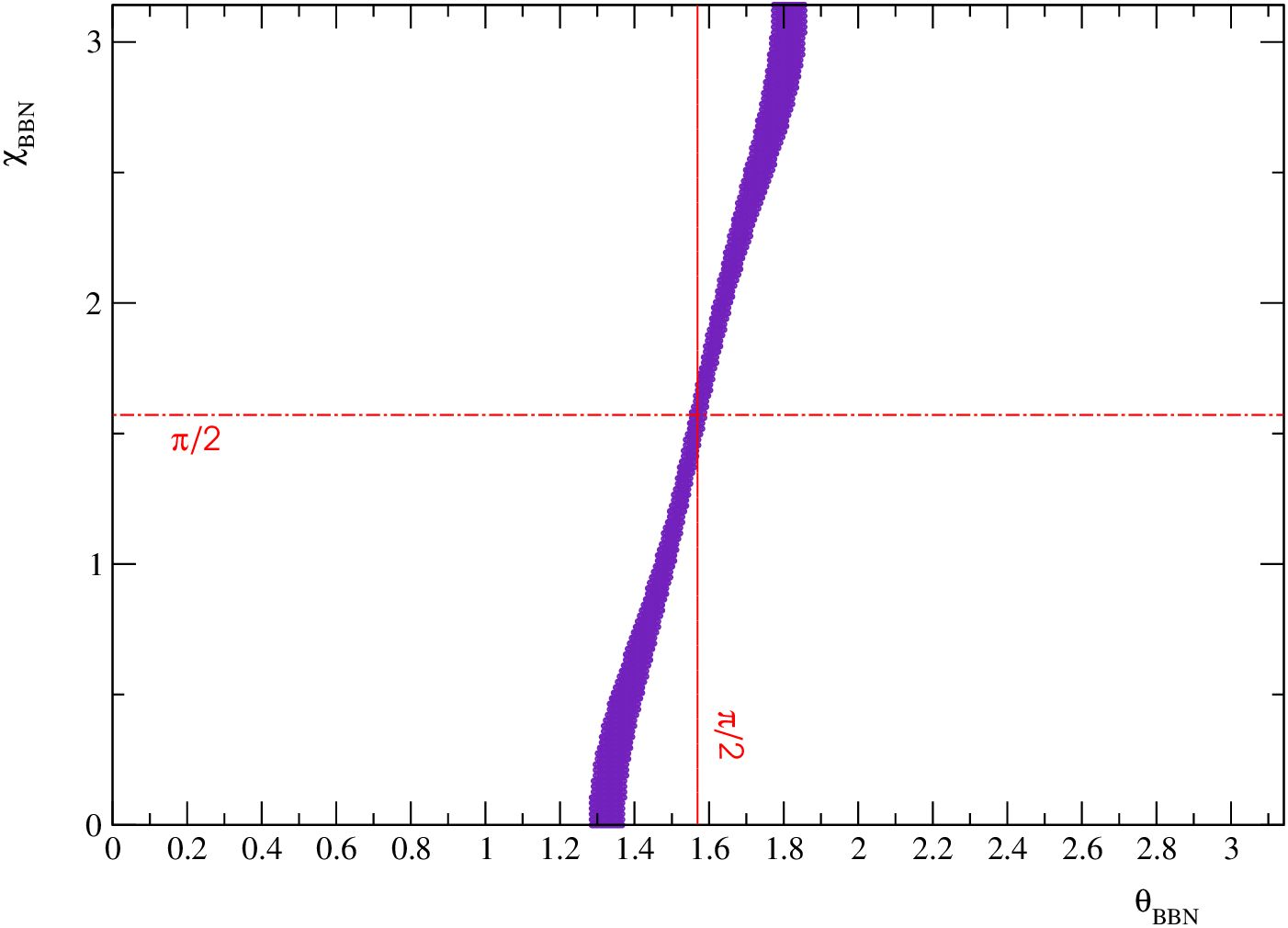}} \caption{The same analysis as for the
previous figure, but with a non-invariant target-space metric
(with symmetry-breaking parameter $c = 10$).} \label{fig:cten}
\end{figure}

\ssubsubsection{Sensitivity to $G$-noninvariant metrics}
We close with a discussion of the sensitivity of the above results
to the choice of an $SO(3)$-invariant target-space metric. To test
this sensitivity, we repeated the above analyses with the
parameter $c$ of eqs.~\pref{SO3breakingterms} nonzero. As
expected, we find that $c$ does not change the scalar cosmology
unless $c$ is quite large. For instance, fig.~\ref{fig:cten} shows
the range of initial conditions which give acceptable present-day
cosmologies if $c = 10$. As is seen from this figure, the allowed
region changes perceptibly relative to the $c=0$ case, but the
total acceptable volume does not change (as might be expected from
Liouville's theorem).

\section{Conclusions}
In this paper we re-examine the cosmological applications of
pseudo-Goldstone bosons, with the following results.

\begin{enumerate}
\item We examine, in $\S2$, the constraints which inflation and
cosmology impose on a slowly-rolling scalar field, and reproduce
there standard constraints which are implied for the scale $f$
associated with the scalar's kinetic energy and the scale $\mu$
related to its potential energy. These typically require $f \gsim
M_p$ and $\mu \ll M_p$.
\item In $\S 3$ we identify a new brane-world mechanism for ensuring
that the scale $\mu$ is exponentially small while keeping $f \gsim
M_p$. It is accomplished by having a theory with an approximate
global symmetry which is broken only by the couplings of a massive
bulk field to various branes. In this model the scale $\mu$ of the
low-energy effective theory below the compactification scale is
proportional to $\exp(-Ma)$, where $M$ is the bulk scalar mass and
$a$ is the inter-brane separation. Once such a small scale is
generated in this way in the low-energy theory it is protected
against low-energy radiative corrections by the residual
approximately-broken symmetry, in the usual manner for a
pseudo-Goldstone boson.
\item Motivated by this mechanism for obtaining extremely small
scales, in $\S 4$ we reconsider the late-time cosmology of such a
pseudo-Goldstone boson, by constructing an explicit quintessence
cosmology. Successful cosmologies can be made subject to mildly
restrictive choices for the initial conditions which are assumed
for the scalars at the epoch of Big Bang nucleosynthesis. We argue
that pseudo-Goldstone bosons of this type will be observationally
distinguishable from other types of quintessence proposals because
of the late-time scalar field oscillations which they generically
predict.
\end{enumerate}

The great difficulty in obtaining slowly-rolling scalar fields
from realistic microscopic theories of physics poses something of
an opportunity given the current observational evidence for two
epochs during which the Universe underwent accelerated expansion.
The challenge is to identify those few kinds of small-distance
physics for which cosmologically acceptable scalar fields are
possible. In the past, brane-world models have been very
successful in circumventing previously-held naturalness obstacles,
and the same may be true for the mechanism illustrated by the
brane-world toy model which we propose here. We believe this class
of models is sufficiently interesting to merit a more detailed
exploration of their observational implications for the CMB.

\section{Acknowledgments}
We would like to acknowledge fruitful discussions with A.
Albrecht, as well as partial research funding from NSERC (Canada),
FCAR (Qu\'ebec) and McGill University. C.B. thanks the KITP in
Santa Barbara for their hospitality while this work was completed
(as such, this research was supported in part by the National
Science Foundation under Grant No. PHY99-07949).

\section{Appendix A: Supersymmetric Models}
In this appendix we briefly summarize how the simple dimensional
estimates of the main text can differ for supersymmetric models.

In $N=1$ supersymmetry in four dimensions scalars arise in complex
pairs, as the partners of spin-1/2 fermions in chiral
supermultiplets. Furthermore, supersymmetry also requires the
quantity $G_{ab}$ can be put into the particular form
\cite{cremmeretal}
\eq \label{E:KahlerPot}
    G_{ab^*} = {\partial^2 K \over \partial \varphi^a \partial
    \varphi^{b*}},
\eeq
for a real function, $K(\varphi,\varphi^*)$, known as the K\"ahler
potential. The scalar potential is similarly given in terms of $K$
and the holomorphic superpotential, $W(\varphi)$ by
\eq \label{E:PfromSP}
    V = e^{K/M_p^2} \, \left[ G^{ab^*} \, D_aW \, (D_b W)^*
     - 3 \, {|W|^2 \over M_p^2} \right],
\eeq
where $D_a W = \partial_a W + \partial_a K \, W/M_p^2$ and
$G^{ab^*}$ denotes the matrix inverse of the target-space metric,
eq.~\pref{E:KahlerPot}.

Additional restrictions arise for $K$ and $W$ if the scalars are
also Goldstone bosons for the symmetry-breaking pattern $G \to H$
\cite{susyGB}. In particular $K$ must be the K\"ahler function for
an appropriate complexification of the manifold $G/H$, and $W$
must be independent of the Goldstone bosons and their
superpartners. For example, for the two-sphere example, $S_2 =
SO(3)/SO(2)$, considered earlier, if the scalars $\theta$ and
$\phi$ are related to one another by supersymmetry, then the
metric has the form of eq.~\pref{E:KahlerPot} when it is expressed
in terms of the stereographic projection to the complex plane,
\eq \label{E:Stereo}
    z(\theta,\phi) = \cot\left({\theta \over 2} \right) \;
    e^{i\phi} \,.
\eeq
The K\"ahler potential in this case is
\eq \label{E:KSphere}
    K(z,z^*) = 4 f^2 \; \log \Bigl(1 + z^* z \Bigr),
\eeq
since with this choice
\eq
    \frac{\partial^2 K}{\partial z \,\partial z^*} \, dz \, dz^* =
    f^2 \, \Bigl( d\theta^2 + \sin^2 \theta \, d\phi^2 \Bigr) \, .
\eeq

For the present purposes, the crucial property of supersymmetric
theories is that $W$ is protected by a nonrenormalization theorem
\cite{susyNRT} and so does not receive corrections to any order in
perturbation theory, although $K$ does. In supersymmetric models
this implies that if a scalar is initially not in the classical
superpotential, it cannot enter in perturbation theory as
successive scales are integrated out, so long as these
integrations remove particles in entire supermultiplets (as is
required if the effective theory is to have the supersymmetric
form given above). If the vacuum is supersymmetric then the
loop-induced dependence of the scalar potential, $V$, on a
pseudo-Goldstone boson must arise through symmetry-breaking
contributions to $K$ rather than $W$.

Once scales of order the supersymmetry-breaking scale, $M_s$, are
integrated out, however, supersymmetry is less restrictive in what
it requires. So if the pseudo-Goldstone boson symmetry-breaking
scale satisfies $\mu \ll M_s$, none of the above discussion is
particularly relevant and the estimates of the main text apply. If
$M_s \ll \mu$, on the other hand, it can happen that corrections
to the K\"ahler function, $K$ --- and so also for the target-space
metric $G_{ab}$, can be larger than those for $V$ if these are
protected by the nonrenormalization theorems.  Consequently
supersymmetric suppressions are not likely to be relevant for
quintessence cosmologies, although they may be relevant for
inflationary models.

\section{Appendix B: Over-Damped, Kinetic-Dominated Scalar Rolls}
In this appendix we identify the $a$ dependence of the scalar
field $\psi$ during a period of kinetic-energy-dominated motion.
In particular, we establish the result $d\psi/da \propto a^{-p}$,
with $p = 3 - m/2$, used in the main text, and derive an upper
limit on the total distance $\psi$ can roll during this kind of
motion.

We start by changing the independent variable from $t$ to $b = \ln
a$, in which case the derivatives of a field $\psi$ become:
\begin{equation}\label{E:ttob1}
\dot{\psi}=\frac{d\psi}{db}\frac{db}{dt}=H\psi'
\end{equation}
\begin{equation}\label{E:ttob2}
\ddot{\psi}=H^{2}\psi''+H'H\psi'
\end{equation}
where over-dots denote $d/dt$ and primes denote $d/db$. If we also
suppose only a single field rolls (so $G_{ab}$ may be set to unity
by performing a field redefinition), then for a kinetic-dominated
roll the Klein-Gordon field equation becomes
\begin{equation}\label{E:roll0}
\psi''+[3+H'/H]\psi'=0 \, .
\end{equation}

Assuming the dominant energy density satisfies $\rho_m \propto
a^{-m}$, with $m = 3$ or 4 for matter- or radiation-domination, we
have $3 M_p^2 H^{2} \approx \rho_m$ and so $H'/H = -m/2$.
Consequently $\psi''+[3-m/2]\psi'=0$, with solution $d\psi/db =
\kappa \, \exp[{-(3-m/2) \, b}]$ with $\kappa$ a constant. Clearly
this establishes $\psi \propto \exp[-(3 - m/2) \, b] \propto
a^{-p}$ with $p = 3-m/2$, as required.

Given this solution we may also compute how far the field rolls,
$\Delta\psi$, in a given amount of universal expansion, with the
result
\begin{eqnarray}\label{E:roll3}
    \Delta\psi & \equiv & \psi_f - \psi_i = \kappa \int_{b_i}^{b_f}
     e^{-(3-n/2)b} \, db \nonumber \\
     & = & \frac{\kappa}{(3-m/2)} \Bigl[  e^{-(3-m/2)b_i} -
       e^{-(3-m/2)b_f} \Bigr] \nonumber \\
     & = & \frac{1}{(3-m/2)} \left[ \left( \frac{d\psi}{db} \right)_{i}
     - \left( \frac{d\psi}{db} \right)_{f} \right] \, .
\end{eqnarray}
We see that $\Delta \psi$ is directly related to the change in the
derivative, $\psi' = (d\psi/db)$, which in turn can be related to
the change in scalar kinetic energy, $K_i = \frac12 \, \dot\psi^2
= \frac12 H^2 \, {\psi'}^2$, between the initial and final times.

Denoting the fraction of energy tied up in the scalar field
kinetic energy by $\varepsilon = K/\rho_m$, we have
\eq
    {\psi'}^2 = \frac{2 K}{H^2} = \frac{6 M_p^2 K}{\rho_m}
    \approx 6 M_p^2 \varepsilon \, .
\eeq
Combining the above results we obtain the final result
\begin{eqnarray}\label{E:roll5}
\frac{\Delta\psi}{M_p} \approx \frac{\sqrt6}{3-m/2} \, \Bigl(
\sqrt{\varepsilon_i} - \sqrt{\varepsilon_f} \Bigr).
\end{eqnarray}
This last expression is useful if the fraction of scalar energy is
known or bounded at the initial and/or final times. For instance,
since constraints from nucleosynthesis require $\epsilon_{BBN}
\lsim 0.1$, this is a useful place to choose as the initial or
final time. A similar observation has also been made in another
context in ref.~\cite{chiba}.

\end{document}